\documentclass[11pt, showpacs, article]{revtex4}

\usepackage{graphicx}%
\usepackage{amsmath}
\usepackage{fancyref}
\usepackage{setspace}
\usepackage[font={small}]{caption}

\newcommand{\mbf}[1]{\mbox{\boldmath $#1$}}

\def\0{\mbox{\mbf 0}}

\def\BibTeX{{\rm B\kern-.05em{\sc i\kern-.025em b}\kern-.08em
    T\kern-.1667em\lower.7ex\hbox{E}\kern-.125emX}}

\begin{document}

\title{Tunneling of electrons via rotor-stator molecular interfaces: combined \emph{ab initio} and model study}

\author{Irina Petreska}
\author{Vladimir Ohanesjan}
\address{ Institute of Physics, Faculty of Natural Sciences and
Mathematics,Ss. Cyril and Methodius University, PO Box 162, 1000
Skopje, Republic of Macedonia, e-mail:irina.petreska@pmf.ukim.mk}
\author{Ljup\v{c}o Pejov}
\address{Institute of Chemistry, Department of Physical
Chemistry, Ss. Cyril and Methodius University, Arhimedova 5, P.O.
Box 162, 1000 Skopje, Republic of Macedonia}
\author{Ljup\v{c}o Kocarev}
\address{Macedonian Academy of Sciences and Arts , Krste
Misirkov 2, PO Box 428, 1000 Skopje, Republic of Macedonia}
\address{Faculty of Computer Science and Engineering, Ss. Cyril and Methodius University, Skopje, Republic of Macedonia}\
\address{BioCircuits Institute, University of
California, San Diego 9500 Gilman Drive, La Jolla, CA 92093-0402 }%


\begin{abstract}

Tunneling of electrons through rotor-stator anthracene aldehyde molecular interfaces is studied with a combined \emph{ab initio} and model approach. Molecular electronic structure calculated from first principles is utilized to model different shapes of tunneling barriers. Together with a rectangular barrier, we also consider a sinusoidal shape that captures the effects of the molecular internal structure more realistically. Quasiclassical approach with the Simmons' formula for current density is implemented. Special attention is paid on conformational dependence of the tunneling current. Our results confirm that the presence of the side aldehyde group enhances the interesting electronic properties of the pure anthracene molecule, making it a bistable system with geometry dependent transport properties. We also investigate the transition voltage and we show that confirmation dependent field emission could be observed in these molecular interfaces at realistically low voltages. The present study accompanies our previous work where we investigated the coherent transport via strongly coupled delocalized orbital by application of Non-equilibrium Green's Function Formalism.

\end{abstract}


\maketitle

\section{Introduction}

Molecular nanostructures, where each individual molecule has its own function are emerging materials for the next generation of electronic devices and components. Due to the increasing success in fabrication of self-assembled molecular monolayers (SAM) and advancement of experimental techniques at one hand, as well as the development of powerful computational facilities for material science simulations on the other, various classes of molecules have been investigated and proved to be promising candidates for future applications \cite{Reed, ref2, Troisi1, Kornilovitch, Troisi2, cunibertibook, ratnernature, lathanature, lathananolett,mitra,13,14,15,16}, and references therein. Ever since the conductance of a molecular junction was experimentally quantified in the mechanically controllable break-junction experiments with benzene 1,4-dithiol SAMs onto golden electrodes \cite{Reed}, benzene-based molecules, also the subject of the present study, have been intensively investigated. Nonlinearity, negative differential resistance, conductance switching, found in various benzene derivatives are some of the many interesting properties that make these systems attractive for investigation from both, experimental and theoretical point of view \cite{ref2,lathananolett,15, 16, maiti, maiti1, nanolett1, ref1, ref3}.  Mechanisms underlying the electron transport in these systems, which are predominated by the quantum-mechanical effects are complex and still a matter of debate. In this respect, theoretical studies and models are crucial for rationalizing the existing experimental data, and even more, for proposing new candidate systems and guiding the fabrication of electronic components.

In the focus of our interest is a class of anthracene based organic molecules containing flexible side groups, denoted as rotor-stator molecular systems \cite{Kornilovitch}. So far, molecular junctions involving anthracene molecules have been extensively investigated. Controllable break junction experiments show that anthracene molecules could be considered as highly conductive single molecule elements, due to direct $\pi$-binding of the molecular orbitals to golden electrodes \cite{nanolett2, liu2015}. Also, clear signatures for the formation of conducting hybrid junctions composed of single anthracene molecules, connected to chains of platinum atoms were recorded by break junction technique in \cite{nanolett1}. This opens the ways towards fabrication and realization of anthracene based molecular electronic components.

Despite of the vast amount of papers dedicated to linear anthracene molecules, studies that are dealing with species containing side groups are lacking. As it was experimentally confirmed that the molecular conductance does not only depend on the chemical properties of the molecule, but it is the conformation that might have a substantial impact on the charge transport \cite{ref2,lathanature, conform}, it should be expected that the presence of side groups in the aromatic compounds will allow some new functions of the future molecular electronic devices. The system we investigate here was proposed in our previous work as a single-molecule switching component, built of anthracene containing aldehyde side group bound to the central benzene ring. We showed that due to the presence of a flexible dipolar aldehyde group, this could act as a bistable rotor-stator system. Depending on the orientation of the side aldehyde group with respect to the three static benzene rings, this system could be found in a stable state, denoted as planar conformer further on, and a metastable state (the top of the double well), when the side group is oriented perpendicularly to the stator part, denoted as perpendicular conformer. Applying the Non-equilibrium Green's Function Formalism (NEGF), we studied the current-voltage ($I-V$) curves in the coherent transport regime, simulating a metal-molecule-metal junction, including also the gating field effects \cite{PPKJCP}. Conformational dependence of the transport properties of such a system was evident, since two distinguishable $I-V$ curves for the two respective conformers were obtained. In \cite{PPKJCP} we also paid a special attention on the possibilities to control the intramolecular torsional dynamics by application of external electric field in the proposed molecule. We also showed that bistability of this system could be controlled and fine tuned by an external electric field, even under realistic conditions, reporting also the temperature dependence of transition probabilities and lifetimes. The previously obtained results, especially the controllability of this molecular device are of crucial importance for further investigations, since stability and control of single-molecule electronic devices is one of the main limitations for operability of the prospective electronic devices at room temperatures.

Transport regime via a metal-molecule-metal junction is in principle dictated by the position of the molecular charge carrying orbitals and electrodes' Fermi level \cite{Nitzan}. Motivated from the most recent findings, regarding the electron tunneling in linear aromatic compounds, showing that there is a gap between the molecular charge carrying orbital and the Fermi level of the metallic electrodes \cite{Reed, liu2015}, in the present study we are focused on the electron transmission via anthracene aldehyde molecular barrier in the tunneling transport regime.  When the molecular charge carrying orbital is above the electrodes Fermi level, than the electron transmission in the metal-molecule-metal junction is seen as tunneling via an insulating barrier. Recently, experimental observations \cite{liu2015, JACS1, JPCC1, JACS2} showed that quasiclassical (WKB) approach gives successful interpretation of the tunneling currents through a molecular insulating barrier and the obtained results fit well with the Simmons' formula \cite{Simmons}. We adopt this methodology for the purposes of our analysis, combining it with accurate electronic structure calculations from first principles. To evaluate the transmission coefficients and tunneling current within the WKB approach, we consider model barriers, encompassing two distinct cases, first neglecting the effects of the molecular internal structure in the transmission process - the case of simple rectangular barrier, and second, we account for these effects via the sinusoidal position dependence of the barrier. We pay special attention of the conformational effects, calculating the current-voltage curves for various orientations of the aldehyde group with respect to the stator part. In the present work we also improve our single molecule analysis, including the end dithiol bonds, that are usually used in fabrication of metal-molecule-metal junctions.

The obtained results for the transmission via the studied molecular segment suggest that molecular internal structure, incorporated in the position dependence of the tunneling barrier along the molecular bridge can not be neglected under realistic conditions. Conformational dependence of the barrier height and thus, the tunneling current confirms that molecular geometry plays crucial role in the electron transport, allowing bistability and enhancing the electronic properties with respect to pure anthracene. Taking into account the recent interest in aromatic compounds, we believe that the proposed model to study tunneling of electrons via molecular interfaces containing side groups, that is complementary to our previous analysis of the coherent transport regime contributes to the modeling and predicting the behavior of newly proposed molecular species, and consequently designing adequate experiments.

The paper is organized as follows: in Section II we introduce our model, in Section III we give an overview of the methodology and computational details, in Section IV results and discussion are presented and Section V comprises our conclusions.

\section{The model}
We consider a model that describes tunneling of electrons in rotor-stator molecular interfaces, incorporating accurate \emph{ab initio} data for anthracene aldehyde molecular species. The model is developed within the frameworks of the WKB approximation, which is valid for barriers having width of the order of 10 angstroms at heights of the order of several eV, applicable in our case \cite{Nitzan}. When one wants to calculate the transmission probability within the WKB approximation, the main problem is to introduce a realistically shaped potential barrier. It is a quite complex task to determine the actual height of the tunneling barrier, taking into account various factors that affect the electronic structure of the molecule when bound to electrodes out of equilibrium. Also the properties of the environment have their impact on the energetic barrier by which the tunneling of electrons in metal-molecule-metal junctions is governed. The simplest, but yet sufficient approach to capture at least the qualitative picture is to consider a rectangular barrier having a height that equals the difference between the Fermi level of the metal and the charge carrying orbital (HOMO for holes and LUMO for electrons). Since we are considering here a newly proposed anthracene based molecular species, we rely on recent results regarding electron tunneling through basic anthracene molecule. A series of experimental works dedicated to anthracene molecule suggest that the Fermi level in the case of dithiol contacts lies closer to the HOMO orbital and it is usually found within the HOMO-LUMO gap \cite{Reed, liu2015, Beebe1, Beebe2}. Having this on mind, we will take the energy of the HOMO orbitals as the first reference point, assuming that the Fermi level is nearly aligned with the HOMO level of the planar conformation. As we already mentioned we are focused on rotor-stator molecular systems from a class of anthracene aldehydes. The presence of a flexible side group bound to the central benzene ring in these systems enables fluctuations of the barrier height due to the changes in the position of frontier molecular orbitals (HOMO and LUMO). Namely, as the rotor part (aldehyde group in this case) undergoes stochastic and thermally activated torsional oscillations the system changes between basically two conformations, planar and a perpendicular one, having different HOMO and LUMO energies. The dependence of the HOMO and LUMO energies on the torsional angle is analyzed in details in our previous work \cite{PPKJCP}. It is evident that fluctuations of the position of HOMO and LUMO orbitals with conformational changes occur. Also, the temperature effects on the transition probability elaborated in our previous work suggest that the tunneling of electrons via such molecules and interfaces would be governed with the geometry dependent potential barrier under realistic conditions. Additionally with the rectangular barrier, we also introduce some more complex sinusoidal position dependence of the tunneling barrier, assuming that the dependence of the frontier orbitals on the geometry of the molecule would result in position dependent tunneling barrier. This position dependence is explained by the effects of the torsional stochastic oscillation of the rotor part, which is situated in the central region, that is not directly attached to the metal leads and thus not significantly affected by the coupling between the molecule and electrodes. Under this circumstances the height of the barrier is derived from \emph{ab initio} calculations for a free molecule, that could be also considered as one building block of a longer molecular wire. In case when the metal Fermi level is aligned with the HOMO of the planar conformer, the barrier height would be expressed as $\varphi=E_i-E_{HOMO,pl}$, where $E_i$ stands for the LUMO energy in each particular conformer. Special attention is paid on the conformational dependence of the tunneling process, so all the calculations are carried out at different position of the rotor part defined by the dihedral angle in the interval from $0^\circ-90^\circ$ by a step of $10^\circ$. Table ~\ref{tab:HLG} lists the variations of the potential barrier heights $\varphi$ with the rotor orientation.
\begin{table}[h]
\begin{tabular}{|l|l|l|l|l|l|l|l|l|l|l|}
\hline
    Angle ($^\circ$) &  0&10&20&30&40&50&60&70&80&90  \\ \hline
   $\varphi \, (\mathrm{eV})$   &  3.15&3.16&3.18&3.23&3.29&3.36&3.44&3.53&3.61&3.64   \\
\hline

\end{tabular}\caption{Geometry dependence of the barrier height. The dihedral angle describes the orientation of the side aldehyde group with respect to stator anthracene structure. Results for each conformer are obtained at B3LYP/LANL2DZ theoretical level, changing the dihedral angle by a step of $10^\circ$.}\label{tab:HLG}
\end{table}

\subsection{Rectangular potential barrier}
As it was mentioned, the potential barrier that would be exerted on incident electrons at the molecule-metal interface is due to energy difference between the metal Fermi level and charge carrying molecular orbitals (LUMO for the case of electrons) \cite{Nitzan, Dong}. As a first assumption we consider tunneling through a rectangular barrier. Delocalization of frontier molecular obitals in the molecule of interest suggests that neglecting the internal molecular structure in shaping the potential barrier would still give a reasonable qualitative description. The width of the potential barrier in our model is approximated with the length of the molecule ($l=1.03 \, \mathrm{nm}$) and the height is determined relative to the electrodes' Fermi level.  When the molecule is biased, the rectangular potential barrier, expressed in electronvolts, assuming that the voltage drop along the molecule is uniform and determined by the applied bias, will now have the following trapezoidal form:
\begin{equation}\label{eq:v1}
V(x)=\left\lbrace
  \begin{array}{c l}
    0,\;\;x\leq0 \;\; \text{and}\;\; x\geq l, & \\ \mu+\varphi-\frac{Ux}{l} ,\;\;0\leq x\leq l,
  \end{array}\right.
\end{equation}
where $\mu$ is the Fermi level of the metal electrode (in our case taken as $5.53 \, \mathrm{eV}$ for gold), $\varphi$ is the barrier height, while $U$ and $x$ are the voltage and the position along the molecule in the direction of electron flow ($x$ in this case), respectively. Under realistic conditions, the Fermi level usually lies within the HOMO-LUMO gap, so another situation would be when the Fermi level lies in the middle of the HOMO-LUMO gap, which is the referent level in the second case. For sake of comparison we also consider this situation as well.

\subsection{Sinusoidal potential barrier}
In the model of rectangular barrier, only the voltage drop along the molecule contributes to position dependence. The second model involves a sine shaped barrier, having maximum value at the central part of the molecule, mimicking the influence of the internal molecular structure on the position dependence of the potential barrier.
In this case we adopt the following barrier shape:
\begin{equation}
V(x)=\left\lbrace
  \begin{array}{c l}
    0,\;\;x\leq0 \;\; \text{and}\;\; x\geq l, & \\ \mu+ \varphi \sin\frac{ \pi x}{l}-\frac{Ux}{l} ,\;\;0\leq x\leq l,
  \end{array}\right.
\label{eq:v2}\end{equation}
where $\varphi$ is again defined as the offset between the corresponding charge carrying orbital and the HOMO orbital of the planar conformation \emph{i.e.} the maximum barrier height, located at the center of the molecule, $l$ is again the total width of the barrier given by the length of the molecule. As the central part is not directly attached to the electrodes the inherent properties of the single molecule would be mostly felt by the transmitted electron at the central part. This is also due to the presence of the rotor part and somewhat more pronounced localization around the central dipolar side group. Another justification for adopting sinusoidal shape as more realistic potential barrier could be also found in the effectively reduced barrier area in this case as a result of the image force.

\section{Methodology and computational details}
\subsection{\emph{Ab initio} electronic structure calculations}
To investigate the single-molecule properties for sulphur-substituted molecule, we employed Hartree-Fock and
density-functional theory (DFT) approach. Within the DFT approach, a combination of the
Becke's three parameter adiabatic connection exchange functional
(B3)\cite{Becke} with the Lee-Yang-Parr correlation one (LYP) \cite{LYP}
was used. The Kohn-Sham SCF equations were iteratively solved using
the LANL2DZ basis set (D95 on first row elements, \cite{LANL2DZ} Los
Alamos effective core potential (ECP) plus DZ on Na-Bi \cite{LANL2DZ1, LANL2DZ2, LANL2DZ3}) for
orbital expansion. The Kohn-Sham (KS) SCF equations were solved
iteratively for each particular purpose of this study, with an
"ultrafine" (99, 590) grid for numerical integration (99 radial and
590 angular integration points). We have shown in our previous
studies \cite{PKCP, PPKPRB} that the performances of B3LYP and B3PW91 levels
of theory for the outlined purposes are essentially identical, while
the HF approach seemed to be unreliable for at least some aspects.
Along with the DFT calculations, we have anyway performed HF SCF
calculations as well, but these were done only for comparison
purposes, and in the paper we will focus solely on the results
obtained at B3LYP/LANL2DZ level of theory. The stationary points on
the potential energy surface (PES) of the studied molecule (Fig. 1)
at the employed theoretical levels were located by Schlegel's
gradient optimization algorithm (employing analytical computation of
the energy derivatives with respect to nuclear
coordinates) \cite{Schlegel}. Their character was further tested by
subsequent computation and analysis of the Hessian matrices. The
absence of negative eigenvalues of the second-derivative matrix
indicated the true minimum character of the particular stationary
point on the PES, while the presence of $n$ negative eigenvalues
indicated that an $n$-th order saddle point is in question. All \emph{ab initio} calculations
were carried out with Gaussian 03 series of computer codes \cite{G03}.

\subsection{Electron transport calculations}

To calculate the tunnelling currents we utilized the method proposed by Simmons \cite{Simmons}. Even though it was initially proposed to describe the tunneling currents through a thin insulating film amid two electrodes, as it was mentioned in the introduction this method has been proved to reproduce equally well some experimentally obtained results of tunneling currents through a single molecule junctions or even to estimate the barrier height from them \cite{JACS1, JACS2}.

The net tunneling current density through the molecule is given by the following equation
\begin{equation}\label{eq:current}
j(U)=\frac{4 \pi m e}{h^3} \int ^{E_m} _0 {T(E_x)\text{d}E_x} \int ^{\infty} _0 {[f(E)-f(E+eU)]\text{d}E_r},
\end{equation}
where $E=E_x + E_r$ is the total electron kinetic energy, considering nearly free electrons in the metal electrode. $E_r$ and $E_x$  are in electrode plane and incident (along tunneling direction)  kinetic energies, respectively. The $T(E_x)$ is the transmission coefficient which gives the tunneling probability of electrons through the molecule. As previously explained we use the WKB approach where the coefficient is calculated with the WKB approximation
\begin{equation}\label{eq:transmission}
T(E_x)=\exp\left[{-\frac{4\pi}{\hbar}\int_{x_1}^{x_2}\sqrt{2m(E_x-V(x))}dx}\right],
\end{equation}
where $V(x)$ is the potential energy barrier along the molecule which obstruct the flow of electrons and $x_1$ and $x_2$ are the classical turning points. It's important to note that the barrier depends on the bias voltage, as shown in (\ref{eq:v1}) and (\ref{eq:v2}).

Working at temperature of $T=0\,\mathrm{K}$, which is a good approximation since any metal electrodes would have much greater Fermi temperature, switching to $\text{eV}$ units and shifting the energy origin at the Fermi level gives following two integrals for the current density
\begin{equation}\label{eq:currentzero}
j(U)=\frac{4 \pi m e^3}{h^3} \left \{ U \int ^{-eU} _{-\mu} {T(E_x) \hspace{2pt} \text{d}  E_x} - \int ^{0} _{-eU} {E_x  T(E_x) \hspace{2pt} \text{d}E_x}  \right \}.
\end{equation}

In this work we do not use the Simmons's approximation using a mean barrier height and neglecting small order terms of the integral (3) in order to obtain analytical expression, known as Simmons' formula. Instead we performed numerical calculations for the transmission coefficient and integrals from the current formula (3). In our model we first use a simple square barrier, using the dimensions of the molecules as obtained from first principles, and then we introduce position dependent barrier having a sine form. The turning points were numerically determined for both cases.

Turning points of the both proposed barrier models and the corresponding transmission coefficients were numerically obtained by using the programming package MATHEMATICA. The transmission coefficients in form of  $[E_x , U]$ matrices were further used to calculate the current densities with the Simpson's quadrature formula \cite{simpson}.

For the $I-V$ plots we present renormalized current density with respect to the maximum value $j_0$, which is obtained for the planar conformation with the sin barrier at the highest voltage used in our calculations ($ U=4\,\text{V}$). In this case cancels all of the constants in front of the integrals in (3) as well as the area through which electrons flow if we calculate the current. Therefore all of the plots present the renormalized currents in the same time eliminating the ambiguous estimation of the tunnelling area.

\section{Results and discussion}
\subsection{Electronic structure}
To analyze the influence of the side dithiol bonds on the electronic structure of the molecule, we have employed several different theoretical levels. In Table ~\ref{tab:electronicstructure} the obtained results from the DFT calculations are compared and we conclude that the HOMO-LUMO gap (HLG) is not significantly affected by the choice of the basis set, which can not be said for the ground-state energies. Also, the comparison with the previously obtained data for the non-substituted species \emph{i. e} without -SH side group shows that the presence of the side -SH groups lowers the HLG for less than $0.2 \, \mathrm{eV}$. This means that calculations for the non-substituted molecule are equally reliable to shape the geometry dependence of the tunneling barrier, because the height of the barrier is primarily determined from the positions of the HOMO and LUMO energies. Along with the DFT calculations, we perform also HF calculations, but as expected the HLG is overestimated (around $8 \, \mathrm{eV}$) within this approximation.
\begin{table}[h]
\begin{tabular}{|l|l|l|l|}
\hline
    Method &   Side atoms       & Ground energy (eV)  & HLG (eV)     \\ \hline
 B3LYP/LANL2DZ & SH fixed      &$-18311.2362448016$  &3.07135        \\ \hline
 B3LYP/6-31+G(d,p)&  SH fixed    & $-39436.40470796$& 3.07053    \\ \hline
 B3LYP/LANL2DZ & SH flexible     &$-18311.21042756$  &3.04550\\ \hline
 B3LYP/6-31+G(d,p)&  SH flexible    & $-39436.34817658$& 3.03842  \\ \hline
 B3LYP/LANL2DZ & no S      &  $-652.745875582$   &3.14695        \\ \hline
 B3LYP/6-31+G(d,p)&  no S    &  $-652.883398864$ & 3.17716  \\ \hline

\end{tabular}\caption{Ground-state energies and the HOMO-LUMO gap (HLG) for the stable planar conformer obtained at several different theoretical levels together with the non-substituted species.}\label{tab:electronicstructure}
\end{table}

\begin{figure}
\center\resizebox{0.8\textwidth}{!}{\includegraphics{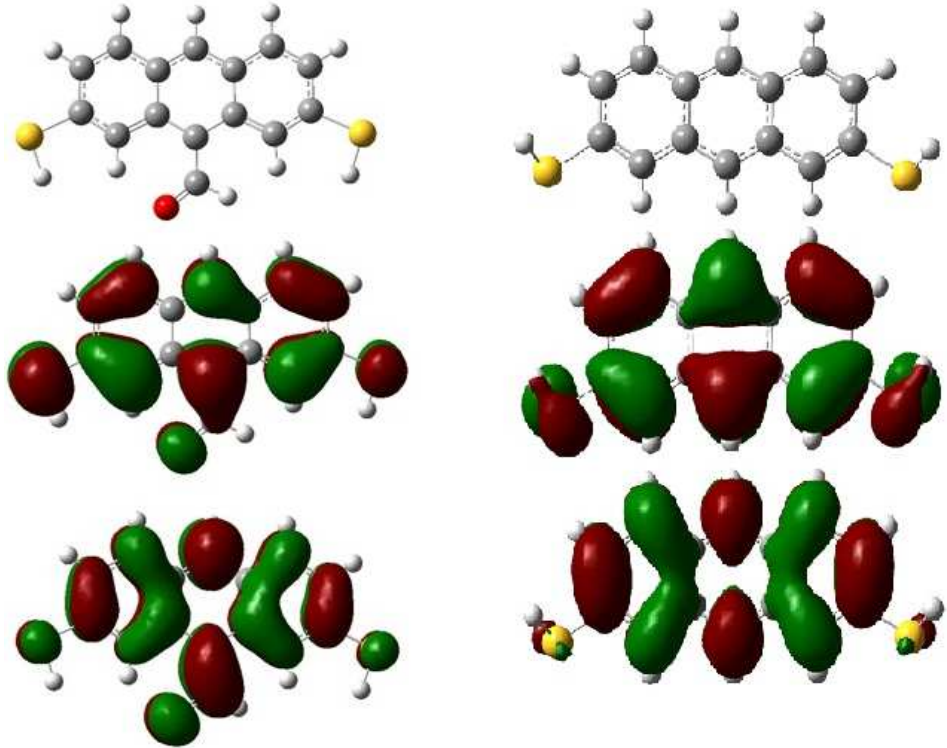}}
\bigskip
\caption{Delocalization of frontier molecular orbitals, HOMO and LUMO for the anthracene aldehyde rotor-stator system (left hand side) together with the corresponding orbitals in anthracene molecule (right hand side). Computations are carried out in presence of dithiol end groups for both species.} \label{fig:Fig.1}
\end{figure}

Geometry dependence of the molecular electronic structure was analyzed by running DFT calculations at different orientations of the side aldehyde group. The obtained results, printed in Table \ref{tab:HLG} show that the dependence of the positions of the frontier molecular orbitals and the HOMO-LUMO gaps on the geometry of the studied conformers should be taken in consideration when modeling the tunneling barrier in rotor-stator molecular species. Delocalization of the charge carrying orbitals of the anthracene aldehyde dithiol molecule in comparison with the basic anthracene is represented in Fig. ~\ref{fig:Fig.1}. It is important to note here that experimental values for the frontier molecular orbitals for a series of organic molecular semiconductors, including anthracene molecule, obtained by inverse photoelectron spectroscopy (IPES) are reported in \cite{djurovich}.

\subsection{Transmission probability and tunneling current}

\begin{figure}
\center\resizebox{0.9\textwidth}{!}{\includegraphics{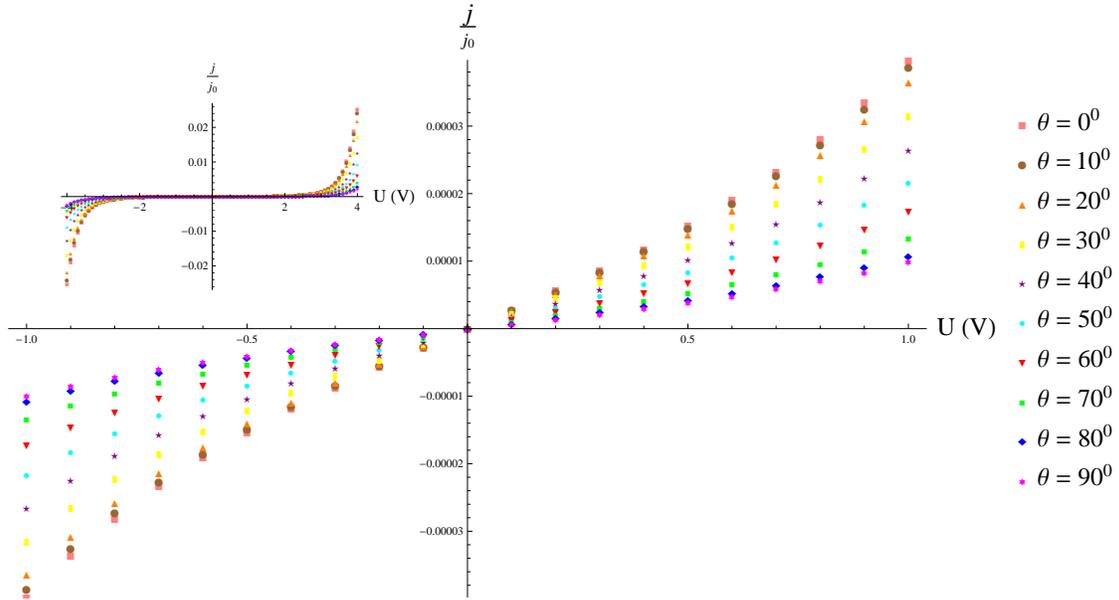}}
\bigskip
\caption{Conformational dependence of the $I-V$ curves for rectangular model barrier. The current density is renormalized with respect to the maximum value $j_0$, which is obtained for the planar conformation with the sine model barrier at the highest voltage used in our calculations ($4\,\mathrm{V}$). The barrier height is approximated with the energy difference between the LUMO orbital in the particular conformation and the HOMO energy of the planar one, while the barrier length is assumed to be equal with the length of the molecule.} \label{fig:Fig.2}
\end{figure}

\begin{figure}
\center\resizebox{0.9\textwidth}{!}{\includegraphics{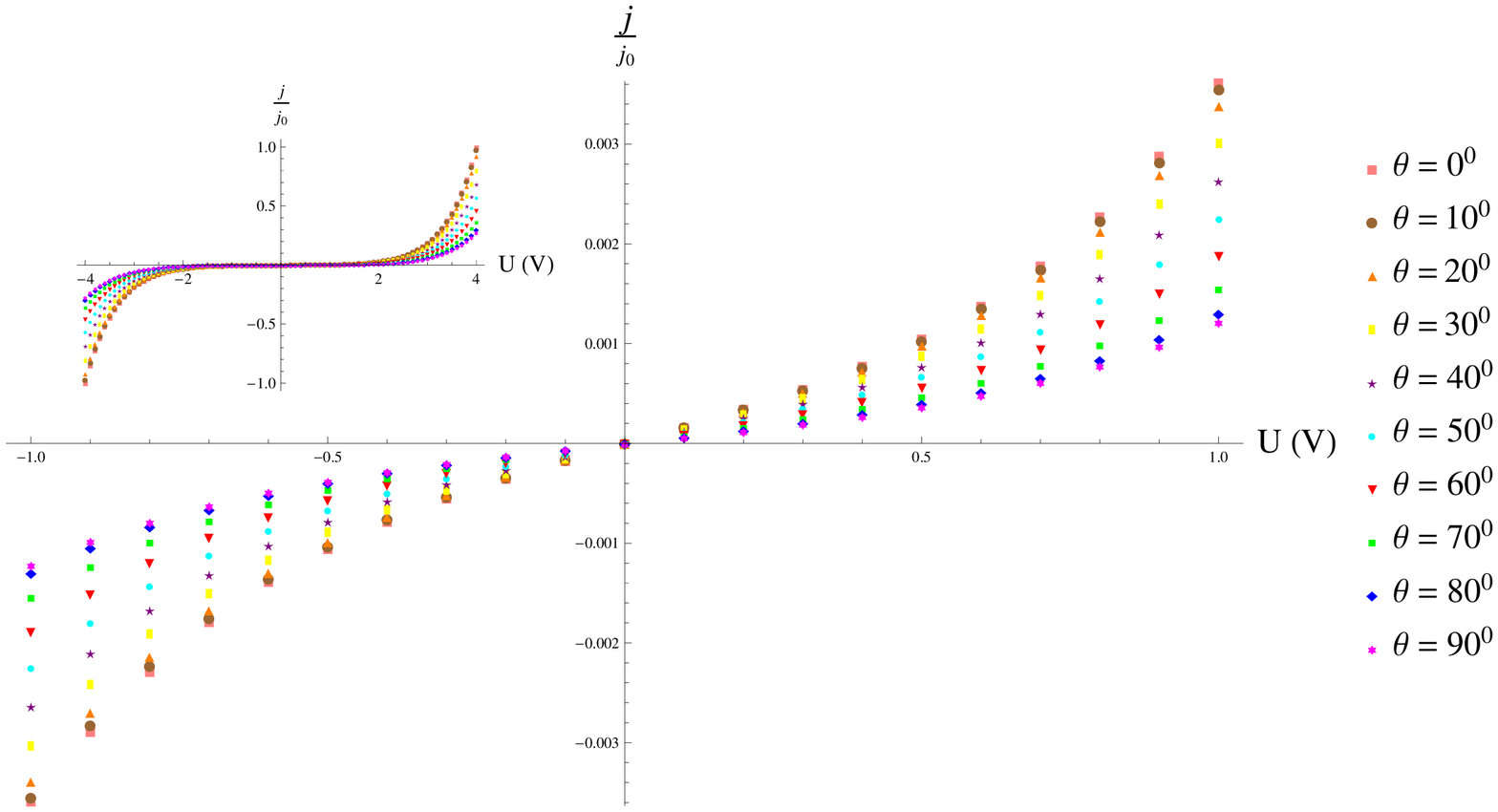}}
\bigskip
\caption{Conformational dependence of the $I-V$ curves for sine model barrier. The current density is renormalized with respect to the maximum value $j_0$, which is obtained for the planar conformation with the sine model barrier at the highest voltage used in our calculations ($4\,\mathrm{V}$). The barrier height is approximated with the energy difference between the LUMO orbital in the particular conformation and the HOMO energy of the planar one, while the barrier length is assumed to be equal with the length of the molecule.} \label{fig:Fig.3}
\end{figure}

Figs. ~\ref{fig:Fig.2} and ~\ref{fig:Fig.3} compare the tunneling currents for square and sine potential barrier, respectively, at different orientations of the side groups. Changes of the dihedral angles from $0^\circ$ to $90^\circ$ lead to variations of the LUMO energies which affects the potential barrier heights. The main plot shows the $I-V$ curves at low voltages, ranging from $-1\,\mathrm{V}$ to $1\,\mathrm{V}$, while the inset of the plot also includes higher voltages in the range from $-4\,\mathrm{V}$ to $4\,\mathrm{V}$. The effects of the side group orientation on the tunneling current are evident and as expected the linear planar conformer has highest conductivity.

Comparison between the models using square potential barrier and sine barrier shows that when one uses a sine position dependence with strong electrode-molecule coupling, which is usually achieved when dithiol covalent bonds are used, higher tunneling current is observed. While if we consider a completely insulating molecular monolayer, taking the whole length of the molecule to model the length of the potential barrier, the obtained current is by several orders of magnitude lower than  the one obtained for sine barrier. However, one should keep on mind that in realistic conditions when the molecule is attached to the leads, the length of the tunneling barrier has somewhat reduced efficient value due to the image force. Calculations of the tunneling via LUMO orbital with the barrier height equal to one half of the HOMO-LUMO gap, which is satisfied when the Fermi level lies in the middle of the molecular energy gap are by several orders of magnitude higher than the ones obtained with the barrier height of the order of the HOMO-LUMO gap. It is important to emphasize that the structural dependence of the tunneling current is evident in all the adopted approaches \emph{i.e.} the fluctuations of the position of the charge carrying orbitals caused by rotation of the side aldehyde group decrease the conductance of the planar conformer by several orders of magnitude with respect to the perpendicular one. The distinguishability between the conductance of the planar and the perpendicular conformer was also confirmed by our previous work where coherent transport regime by applying NEGF formalism was studied \cite{PPKJCP}. Analysis of the tunneling current in the present work, carried out with several approaches using the values of the potential barrier height obtained by accurate \emph{ab initio} calculations accomplishes our previous work confirming that such rotor-stator molecular systems could be implemented as field-controllable single-molecule switches. Under realistic conditions when the torsional oscillations are strongly thermally activated and current fluctuations would be observed in this single-molecule device. To avoid such fluctuations, as it was shown in our previous work external electric field could be used to stabilize even the metastable perpendicular state.

\begin{figure}
\center\resizebox{0.9\textwidth}{!}{\includegraphics{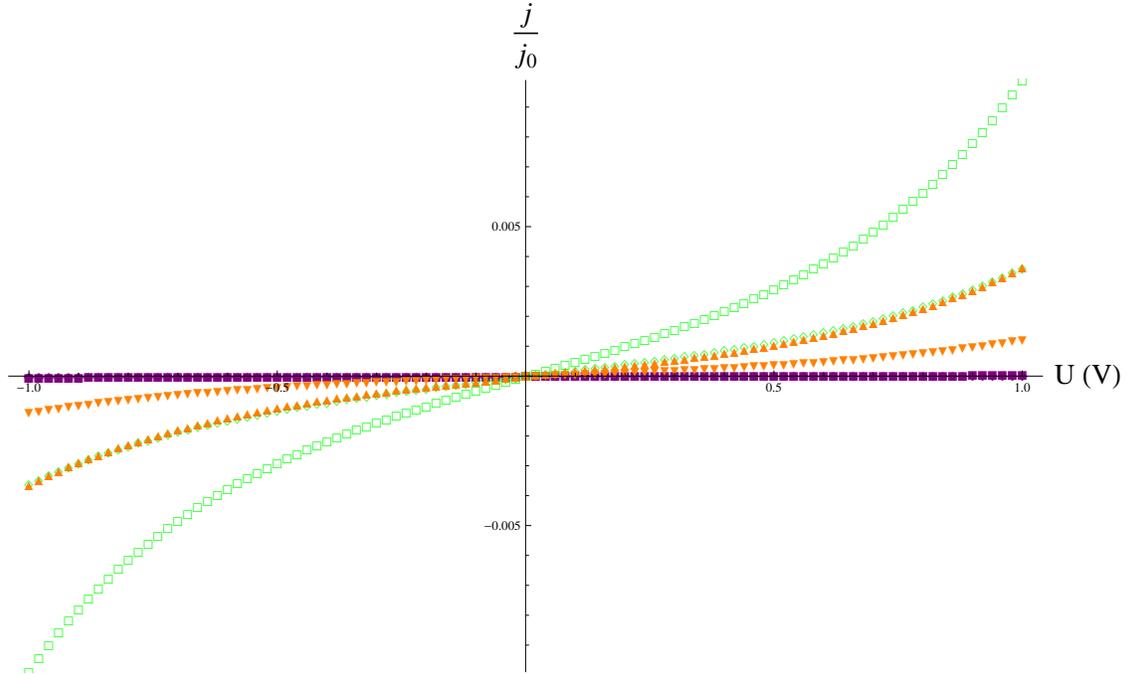}}
\bigskip
\caption{Comparison between currents at low voltages for planar and perpendicular conformers between the cases when the barrier height is taken as a full difference between the LUMO and the HOMO of the planar conformer and as a half this difference. Purple markers denote square barrier with full difference (Fermi level is aligned with the HOMO energy of the planar conformer), green markers stand for square barrier with half height (Fermi level is positioned in the middle of the HOMO-LUMO gap), orange markers are for sine barrier. The current density is renormalized with respect to the maximum value $j_0$, which is obtained for the planar conformation with the sine model barrier at the highest voltage used in our calculations ($4\,\mathrm{V}$).The barrier width is assumed to be equal with the length of the molecule.} \label{fig:Fig.4}
\end{figure}

\begin{figure}
\center\resizebox{0.9\textwidth}{!}{\includegraphics{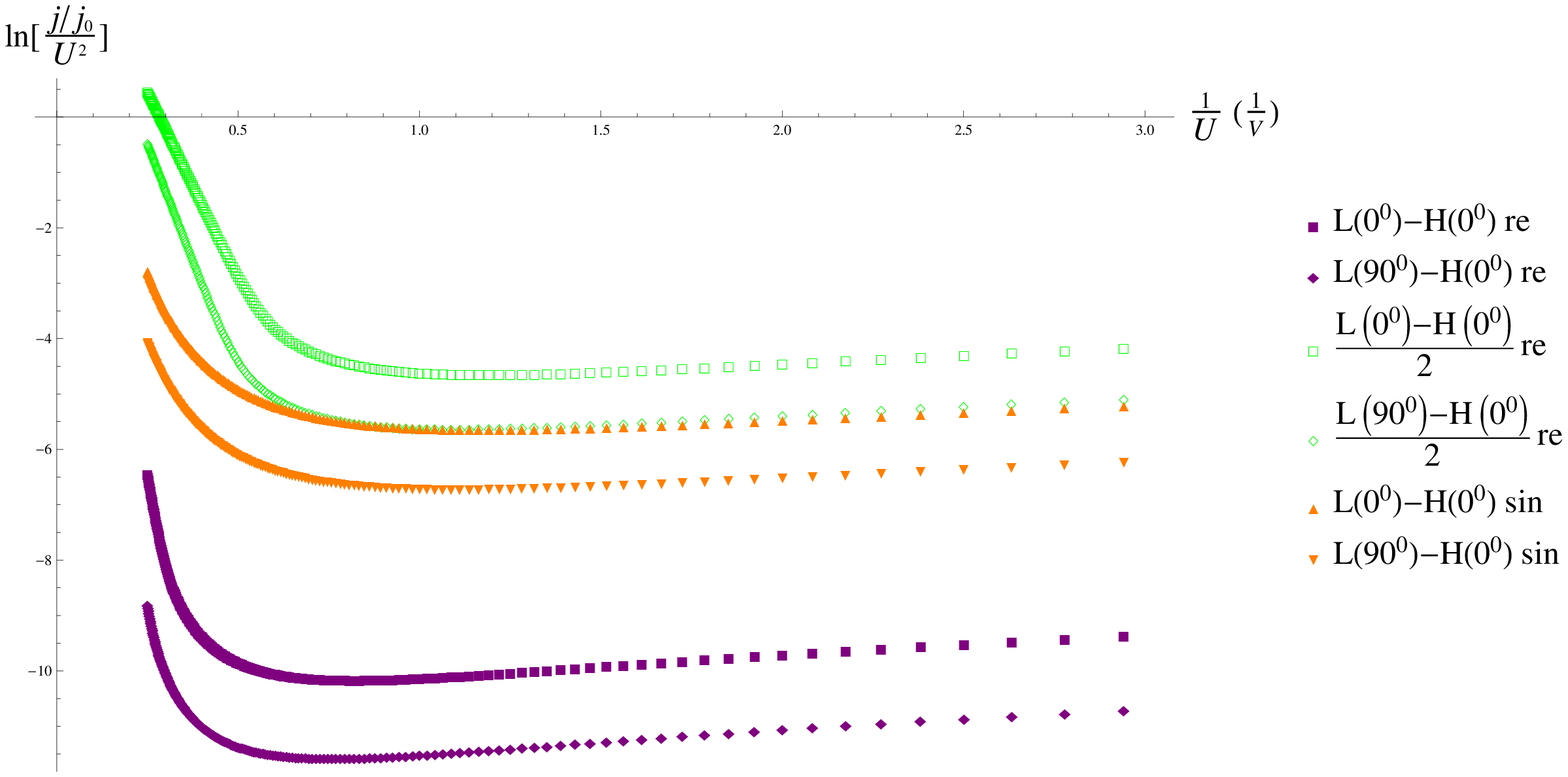}}
\bigskip
\caption{Conformational dependence of the transition voltage curves. The current density is renormalized with respect to the maximum value $j_0$, which is obtained for the planar conformation with the sine model barrier at the highest voltage used in our calculations ($4\,\mathrm{V}$). As in the previous figure, three distinct cases are represented here, rectangular barrier where the Fermi level is aligned with the HOMO energy of the planar conformer, rectangular barrier where the Fermi level is positioned in the middle of the HOMO-LUMO gap and the sinusoidal barrier having maximum height at the center of the molecule. For the barrier length we assume to be equal to the length of the molecule.} \label{fig:Fig.5}
\end{figure}

In order to test the relevancy of the proposed potential barrier and to inspect transition between direct tunneling to field emission, we adopt the approach from the recent work by Beebe et al. \cite{Beebe1, Beebe2} and we plot $ln(I/V^2)$ vs $1/V$. The inflection points of these curves is used to determine the value of the voltage at which the transition from direct tunneling to field emission occurs. If one compares the values that can be read directly form the plot in Fig. \ref{fig:Fig.4} for several models of the tunneling barrier, considered in the present work with the values obtained for the basic anthracene by Beebe et al. with the transition voltage spectroscopy (TVS) higher transition voltage in anthracene with side group would be found. The TVS measurements lead to the value of $0.6\,\mathrm{V}$ for the transition voltage value in the afore cited work, while the value obtained for our rotor-stator as can be seen from Fig. \ref{fig:Fig.5} system is around $1\,\mathrm{V}$  for the planar conformation. As expected the inflection point is shifted towards lower voltages in the model with lower barrier height. Also, the conformational dependence affects the transition voltage which could be concluded from the shift of the curve for perpendicular conformer towards higher voltages with respect to the planar one. The transition voltage  obtained here would be probably overestimated since the approximation of the barrier height with the total value of the HOMO-LUMO gap for a free molecule is certainly higher than the actual HOMO-LUMO gap of the molecule bound to electrodes. Referring to the electric field effects on the HOMO-LUMO gap, obtained in our previous work one could claim that the biasing field applied along the major molecular axis, parallel to the plane of anthracene fragment would decrease the HOMO-LUMO gap (see Fig. 7 in \cite{PPKJCP}). This means that the barrier height is expected to be lower than the one consider in our model and thus, the field emission in these metal-molecule-metal junction should be possible at realistically low voltages.

%

%

\section{Conclusions}
In the present paper bridge assisted electron transfer via rotor-stator molecules or molecular interfaces is considered. Particularly, tunneling processes through anthracene aldehyde species are investigated. Bistability and conformation dependent electron transfer make these species interesting for molecular electronics. WKB approach and Simmons' formula are applied to calculate the transmission probability and tunneling current, proposing model barriers derived from \emph{ab initio} electronic structure calculations. Special attention is paid on the dependence of the tunneling current on the molecular geometry. For this purpose a series of electronic structure calculations are carried out at various orientations of the side aldehyde group. To evaluate the tunneling current we consider basically two different model barriers, a rectangular and a sinusoidal barrier. The barrier heights are determined from first principles considering the variations of the HOMO-LUMO gap with the torsional rotation of the rotor part. By considering a sinusoidal barrier we also incorporate the position dependence of the tunneling barrier that is due to the molecular internal structure together with the usual voltage drop arising from the external bias. Our results suggest that tunneling processes in this species are strongly dependent on the molecular conformation and the molecular internal structure must be taken into account to derive a realistic model for describing such processes. The fluctuations of the charge carrying orbitals, due to torsional stochastic oscillations of the rotor part of the molecule significantly affect the tunneling current.  Accounting for the effects of nuclear motion is required by the fact that structural stability of such systems is crucial for prospective practical applications. The present study accompanies our previous work where the coherent transport by application of NEGF formalism is investigated. Within both transport regimes, coherent via strongly coupled delocalized orbital or tunneling via an insulating bridge, a clear difference between the planar and perpendicular molecular conformation is recorded. This confirms that the presence of a side aldehyde group that turns the simple anthracene to a rotor-stator system enriches the interesting and promising properties of the anthracene based molecules, widening the possible areas of application.

\end{document}